\documentclass{iau} 
\usepackage{graphicx}

\title[Magnetic fields of (post-)AGB stars] 
{Magnetic fields of AGB and post-AGB stars}

\author[Wouter Vlemmings]   
{Wouter Vlemmings$^1$}

\affiliation{$^1$Chalmers University of Technology, Department of
  Space, Earth and Environment, \\ Onsala Space Observatory,
  SE-43992 Onsala, Sweden \\ email: {\tt wouter.vlemmings@chalmers.se}}

\pubyear{2018}
\volume{S343}  
\jname{Title of your IAU Symposium}
\editors{A.C. Editor, B.D. Editor \& C.E. Editor, eds.}
\begin{document}

\maketitle

\begin{abstract}
  There is ample evidence for strong magnetic fields in the envelopes of
  (Post-)Asymptotic Giant Branch (AGB) stars as well as supergiant stars. The origin and role of
  these fields are still unclear. This paper updates the current
  status of magnetic field observations around AGB, post-AGB stars
  and describes their possible role during these stages of
  evolution. The discovery of magnetically aligned dust around a
  supergiant star is also highlighted. In our search for the origin of the magnetic fields,
  recent observations show the signatures of possible magnetic
  activity and rotation, indicating that the magnetic fields might be
  intrinsic to the AGB stars.  
\end{abstract}

\firstsection 
\section{Introduction}

Magnetic fields are ubiquitous throughout the Universe and play an
important role across a wide range of scales. Primordial magnetic
fields could have played a role in the formation of the first stars
just as magnetic fields in molecular clouds are an important
ingredient in current star formation. Magnetic fields have also been
detected in almost all stellar types and in almost all phases of
stellar evolution (e.g. Berdyugina 2009), and have significant effects
on stellar evolution through, e.g. their influence on the internal
mixing.  The magnetic field of stars can have either a dynamo origin,
i.e. be generated by a dynamo process in the star itself
(e.g. Charbonneau 2014), or can be the result of a remnant 'fossil'
field, which are fields that originate from the star formation process
(e.g. Braithwaite \& Spruit 2004). The stellar magnetic field is
affected by the changes of physical properties during stellar
evolution and, because of flux conservation, becomes increasingly
difficult to observe at the stellar surface when the star expands in
the final phases of its life. However, in stellar end products, such
as white dwarfs and neutron stars, magnetic fields are also shown to
be significant.

The role of magnetic fields around AGB stars is not clear. In
principle, they could help levitate material off the stellar surface
through Alfv{'e}n waves (e.g. Falceta-Gon{\c c}alves \& Jatenco-Pereira
2002), or through the creation of cool spots on the surface above with
dust can form easier (Soker 1998). A specific model for the AGB star o
Ceti (Mira A) has shown that a hybrid magnetohydrodynamic-dust-driven
wind scenario can explain its mass loss (Thirumalai \& Heyl 2013). In
such a model, Alfv{\'e}n waves add energy to lift material before dust
forms and radiation pressure accelerates a wind. Magnetic fields also
play an important role in the internal mixing required for s-process
(slow) neutron capture reactions that define the stellar yields
(e.g. Trippella et al. 2016).

After the AGB phase, the stellar envelopes undergo a major
modification as they evolve to Planetary Nebulae (PNe). The standard
assumption is that the initial slow AGB mass loss quickly changes into
a fast superwind, generating shocks and accelerating the surrounding
envelope (Kwok et al. 1978). It is during this phase that the
typically spherical CSE evolves into a Planetary Nebula. As the
majority of of pre-PNe are aspherical, an additional mechanism is
needed to explain the departure from sphericity. This mechanism is
still a matter of fierce debate. One possibility is that the
interaction of the post-AGB star and a binary companion or massive
planet supports a strong magnetic field that is capable of shaping the
outflow (e.g. Nordhaus et al. 2007).

This paper expands on (and partly reproduces) the reviews presented in
\cite[Vlemmings 2018]{Vlemmings18a} and \cite[Vlemmings
2014]{Vlemmings14} and I refer interested readers to
those review (and references therein) for further background. 

\section{Overview of magnetic field observations}

\begin{figure}
\begin{center}
\includegraphics[width=10cm]{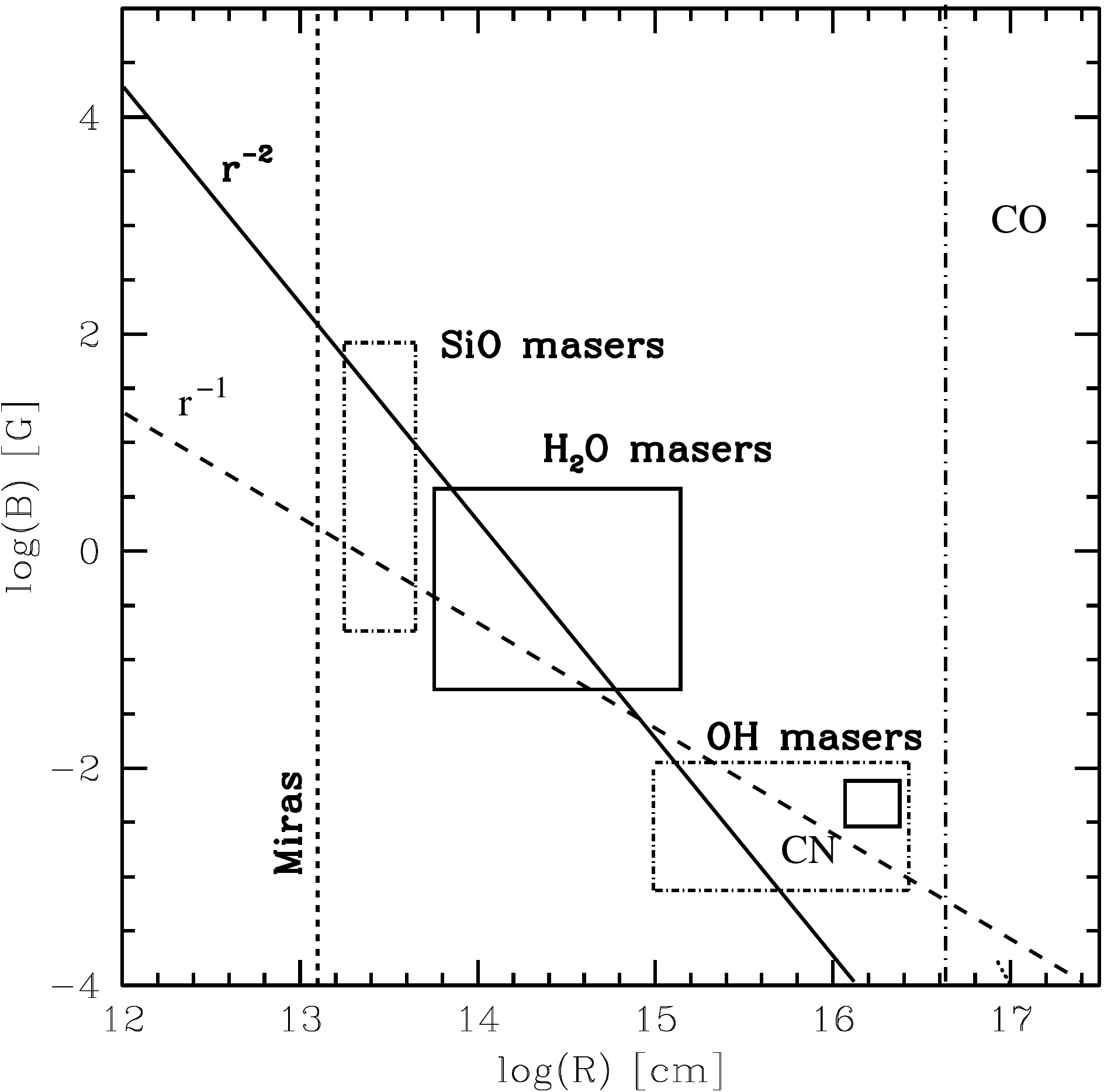} 
\caption{The circumstellar magnetic field strength vs. radius relation as indicated by current (maser)
polarization observation. The boxes show
the range of observed magnetic field strengths derived from the
observations of SiO masers (\cite[Kemball et
al. 2009]{Kemball2009}, \cite[Herpin et al. 2006]{Herpin2006}), H$_2$O masers
(\cite[Vlemmings et al. 2002]{Vlemmings2002},\cite[Vlemmings et
al. 2005]{Vlemmings2005}, \cite[Leal-Ferreira et al. 2013]{Ferreira2013}), OH masers
(\cite[Rudnitski et al. 2010]{Rudnitski2010}, \cite[Gonidakis et
al. 2014]{Gonidakis2014}) and CN (\cite[Duthu et al. 2017]{Duthu2017}). The thick solid
and dashed lines indicate an $r^{-2}$ solar-type and $r^{-1}$
toroidal magnetic field configuration. The vertical dashed line
indicates the stellar surface. Observations of the Goldreich-Kylafis
effect in CO (\cite[e.g. Vlemmings et al. 2012]{Vlemmings2012}) will
uniquely probe the outer edge of the envelope (vertical dashed dotted
line).}
\label{fig1}
\end{center}
\end{figure}

\subsection{AGB stars}

Generally, AGB magnetic field measurements come from maser
polarization observations (SiO, H$_2$O and OH). These have revealed a
strong magnetic field throughout the circumstellar
envelope. Figure~\ref{fig1}, the magnetic field strength in the
regions of the envelope traced by the maser measurements throughout
AGB envelopes. The field appears to vary between $B\propto R^{-2}$
(solar-type) and $B\propto R^{-1}$ (toroidal). Although the maser
observations trace only oxygen-rich AGB stars, recent CN Zeeman
splitting observations (\cite[Duthu et al. 2017]{Duthu2017}) indicate
that similar strength fields are found around carbon-rich stars. The
envelope magnetic fields are also consistent with thus far the only
direct measurement of the Zeeman effect on the surface of an AGB star,
the Mira variable star $\chi$~Cyg (\cite[L{\`e}bre et
al. 2014]{Lebre2014}).  In Table.~1 an overview is given of the energy
densities throughout the AGB envelopes.

The large-scale structure of the magnetic field is more difficult to
infer, predominantly because the maser observations often probe only
limited line-of-sights. Even though specifically OH observations seem
to indicate a systematic field structure, it has often been suggested
that there might not be a large-scale component to the field that
would be necessary to shape the outflow (Soker 2002). Until recently
the only tight shape constraints throughout the envelope had been
determined for the field around the supergiant star VX Sgr, where
maser observations spanning 3 orders of magnitude in distance are all
consistent with a large scale, possibly dipole shaped, magnetic field
(Vlemmings et al. 2005, Vlemmings et al. 2011).

\begin{figure}
\centering
\includegraphics[width=13.5cm]{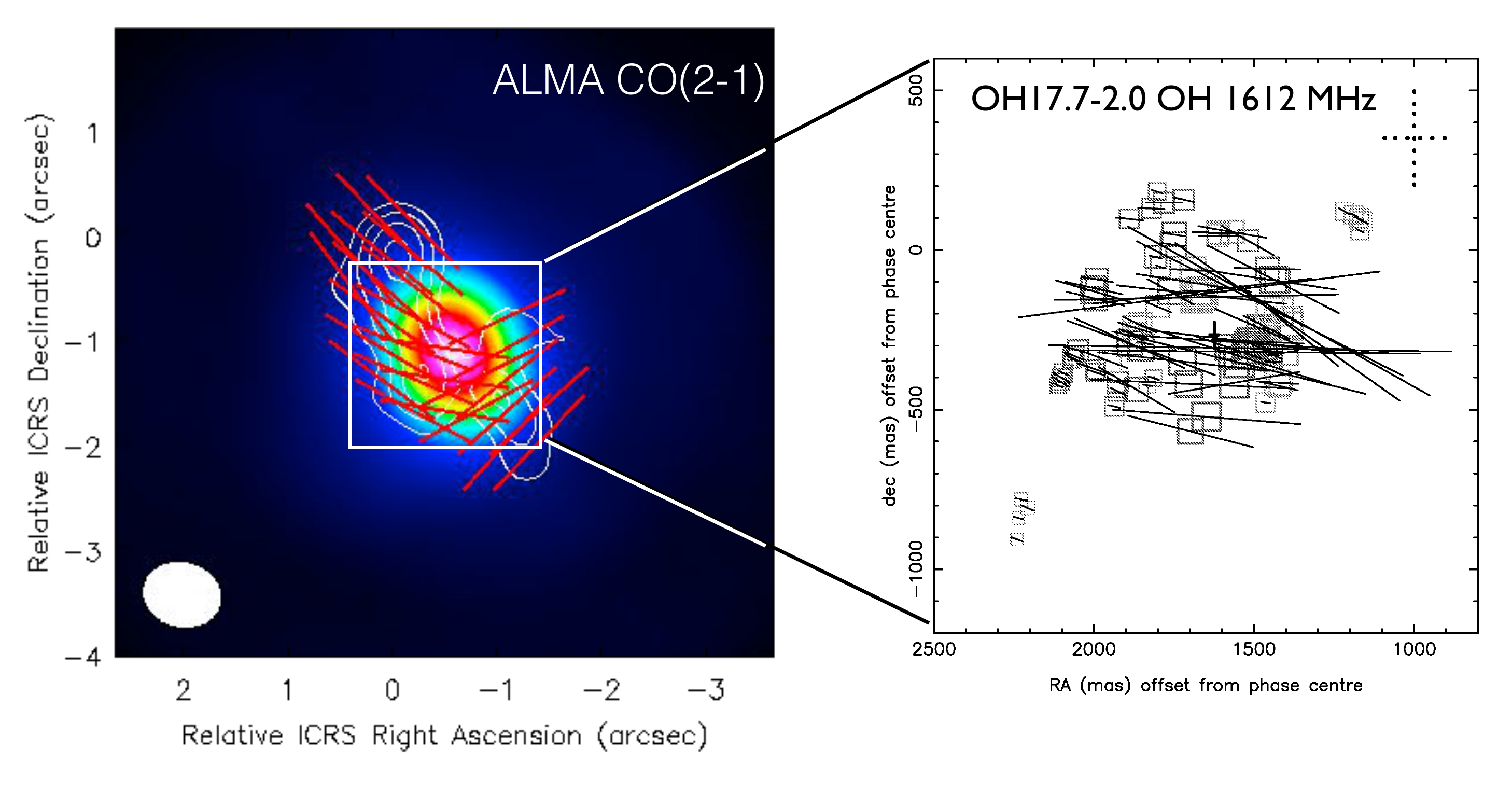}
\caption{A comparison of the magnetic field determined using ALMA
  observations of the Goldreich-Kylafis effect on circumstellar CO
  (left, Tafoya \& Vlemmings, in prep.) and MERLIN observations of OH masers (right, Bains et al. 2003) around the post-AGB star OH 17.7-2.0. These observations show that the CO and OH trace the same large-scale magnetic field.}
 \label{OH17}
\end{figure}

\begin{figure}
\centering
\includegraphics[width=13.5cm]{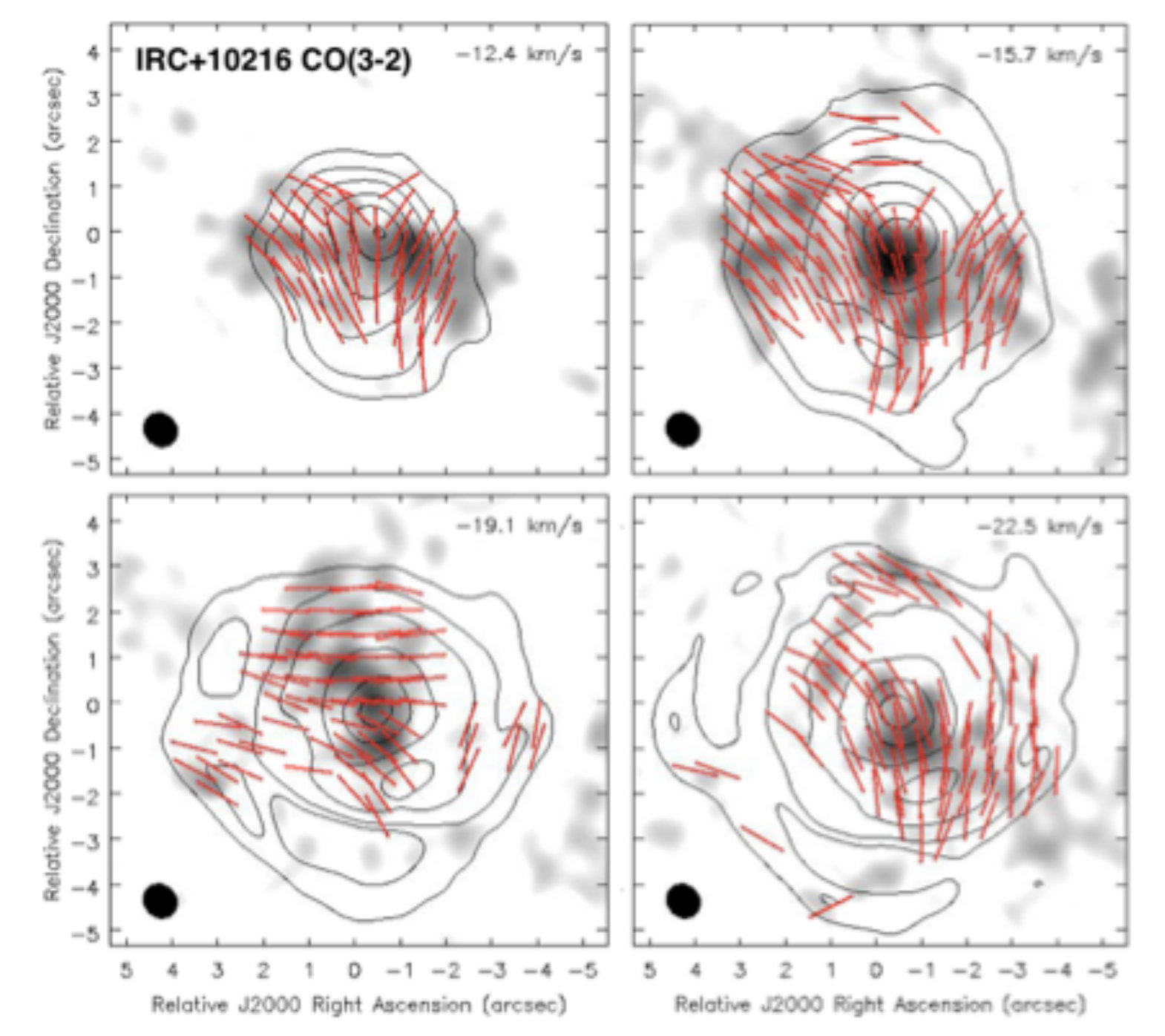}
\caption{Four channel maps showing the Goldreich-Kylafis effect on the
  CO(3-2) line in the envelope of the AGB star IRC+10216. The ALMA
  observations for the first time clearly resolve the magnetic field
  structure throughout the CSE. The red vectors indicate the
  polarisation direction. The linearly polarised emission is shown in
  greyscale. The contours indicate the total intensity emission. A
  structure function analysis will be able to reveal the velocity
  resolved field strength and initial indication indicate a structured
  field with a strength $>1$~G at the stellar surface(Vlemmings et al., in prep.).}
 \label{IRC}
\end{figure}

Very recent ALMA observations have shown that it will soon be possible
to finally overcome the problems with determining the circumstellar
magnetic field structure. This involves observations aimed at
measuring the Goldreich-Kylafis effect, which allows us to use the
polarisation of non-maser molecular lines (in this case CO) to
determine the magnetic field morphology in the more diffuse
circumstellar gas. The first of these observations, for the post-AGB
star OH~17.7-2.0, indicate that the magnetic field structure probed by
the CO is consistent with that derived from OH maser observations
(Fig.~\ref{OH17}, Tafoya \& Vlemmings in prep.). This puts to rest the
decades old question if maser magnetic field measurements can really
be used to probe the large-scale fields. The second set of
observations has given us the first view velocity resolved view of
the large-scale magnetic field in the AGB stars IRC+10216
(Fig.~\ref{IRC}, Vlemmings et al. in prep.).

\subsection{post-AGB stars}

Similar to the AGB stars, masers are the main source of magnetic field
information of post-AGB and P-PNe and even for some PNe. OH maser
observations indicate magnetic field strengths similar to those of AGB
stars (few mG) and a clear large scale magnetic field structure
(\cite[Bains et al. 2003]{Bains2003}, \cite[G{\'o}mez et
al. 2016]{Gomez2016}). Also dust polarization observations indicate a
large scale magnetic field (\cite[e.g. Sabin et
al. 2015]{Sabin2015b}).

Magnetic fields have also been detected around the so-called
'water-fountain' sources. These sources exhibit fast and highly
collimated H$_2$O maser jets that often extend beyond even the regular
OH maser shell. With the dynamical age of the jet of order 100 years,
they potentially are the progenitors of the bipolar
(P-)PNe. Observations of the arch-type of the water-fountains, W43A,
have revealed a strong toroidal magnetic field that is collimating the
jet (\cite[Vlemmings et al. 2006]{Vlemmings2006}). For another water-fountain source,
IRAS~15445-5449, a synchrotron jet related to strong magnetic fields
has been detected (\cite[P{\'e}rez-S{\'a}nchez et al. 2013]{PerezSanchez2013a}). Similar, synchrotron
emission has been found from what could be one of the youngest PNe
(\cite[Su{\'a}rez et al. 2015]{Suarez2015}).

Finally, recently also surface fields have been measured for 2
post-AGB stars (\cite[Sabin et al. 2015]{Sabin2015a}). These fields are consistent with
the fields inferred from the envelope measurements

\begin{center}
\begin{table}
\caption{Energy densities in AGB envelopes}
\begin{tabular}{ l l || c | c | c | c | c }
\hline
 & & Photosphere & SiO & H$_2$O & OH & CO/CN \\
\hline
 & & & & & & \\
 $B$ & [G] & $\sim1-10$? & $\sim3.5$ & $\sim0.3$ & $\sim0.003$ & $\sim0.003-0.008$\\
 $R$ & [AU] & - & $\sim3$ & $\sim25$ & $\sim50$ & $\sim50-100$ \\
$V_{\rm exp}$ & [km~s$^{-1}$] & $\sim20$ & $~\sim5$ & $\sim8$ &
$\sim10$ & $\sim10$ \\
 $n_{\rm H_2}$ & [cm$^{-3}$] & $\sim10^{11}$ & $\sim10^{10}$&
 $\sim10^{8}$ & $\sim10^{6}$ & $\sim10^{5}$ \\
 $T$ & [K] & $\sim2500$ & $\sim1300$ & $\sim500$ & $\sim300$ & $\sim150$\\
 & & & & & \\
\hline 
 & & & & & \\
 $B^2/8\pi$ & [dyne~cm$^{-2}$] & $\mathbf{10^{-1.4,+0.6}}$ &
 $\mathbf{10^{+0.1}}$ & $\mathbf{10^{-2.4}}$ & $10^{-6.4}$ & $\mathbf{10^{-6.0,-6.4}}$\\
 $nKT$ & [dyne~cm$^{-2}$] & $10^{-1.5}$ & $10^{-2.7}$ & $10^{-5.2}$ &
 $10^{-7.4}$ &  $10^{-8.7}$\\
 $\rho V_{\rm exp}^2$ & [dyne~cm$^{-2}$] & $10^{-0.3}$ & $10^{-2.5}$ &
 $10^{-4.1}$ & $\mathbf{10^{-5.9}}$ & $10^{-6.9}$ \\
 $V_A$ & [km~s$^{-1}$] & $\sim20$ & $\sim100$ & $\sim300$ & $\sim8$&
 $\sim8$ \\
 & & & & & \\
\hline
\end{tabular}
\label{energy}
\caption{Energy densities through AGB star envelopes. From left to
  right the columns indicate the stellar photosphere, maser regions
  and the region probed by CO/CN, with increasing distance to the
  central star. The top rows are the typical magnetic field strength
  $B$, distance to the star $R$, expansion velocity $V_{\rm exp}$,
  hydrogen number density $n_{\rm H}$ and temperature $T$. The bottom
  rows are the magnetic, thermal and kinematic energy and a rough
  estimate of the Alvf{\'e}n velocity $V_A$.}
\end{table}
\end{center}

\subsection{Supergiant stars}

\begin{figure*}
 \centering
 \includegraphics[width=12cm]{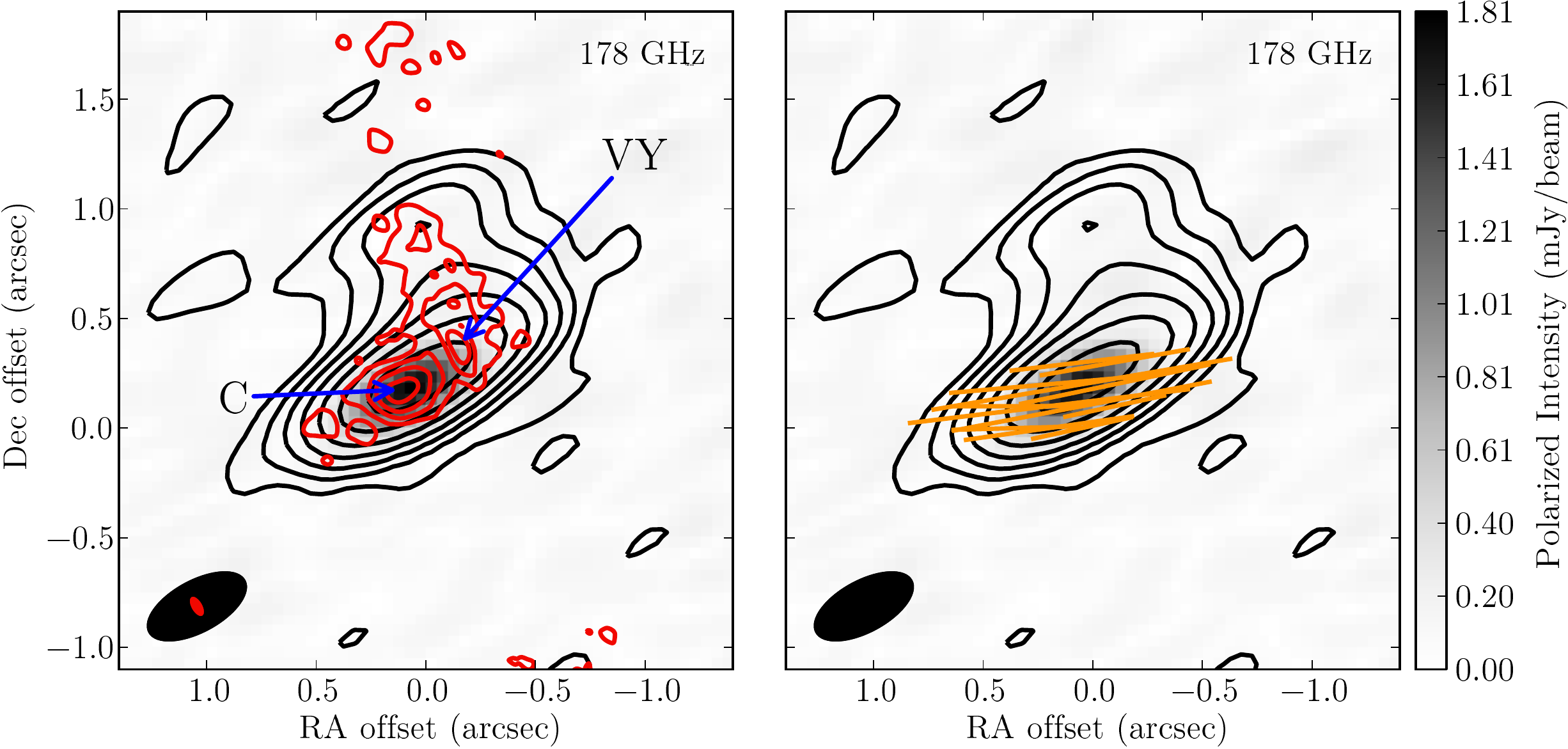}
 \caption{ALMA observations of the dust around the RSG VY~CMa at
   178~GHz (\cite[Vlemmings et al. 2017]{Vlemmings2017a}). Arrows
   indicate a strong dust clump (C)
   and the star (VY). The grey scale image
   is the linearly polarized intensity. The similarly spaced red contours
   (left) indicate the ALMA 658~GHz continuum from \cite[O'Gorman et
   al. 2015]{OGorman2015}. The vectors
   (right) indicate the direction of the magnetic field traced by
   magnetically aligned dust grains.}
       \label{VY}
 \end{figure*}

Many maser observations show that strong magnetic fields are also
present in the envelopes of Red Supergiant stars (\cite[e.g. Vlemmings
et al. 2002, Herpin et al. 2006]{Vlemmings2002, Herpin2006}). The
questions about local or large scale fields, are the same as around
AGB stars. As noted above, the supergiant VX~Sgr is one of the first
stars where a large scale magnetic field, with a structure consistent
throughout the envelope, was found. At (sub-)millimeter wavelengths it
is now possible to simultaneously study the polarization of masers,
regular molecular lines, and circumstellar dust using ALMA. Recent observations
of VY~CMa indicate magnetically aligned dust and consistent structures
between the maser and non-maser molecular lines (\cite[Fig.~\ref{VY},
Vlemmings et al. 2017]{Vlemmings2017a}). The observations indicate that
magnetic fields could be involved in the mass loss of these massive
stars.

\begin{figure}
\begin{center}
\includegraphics[width=13.5cm]{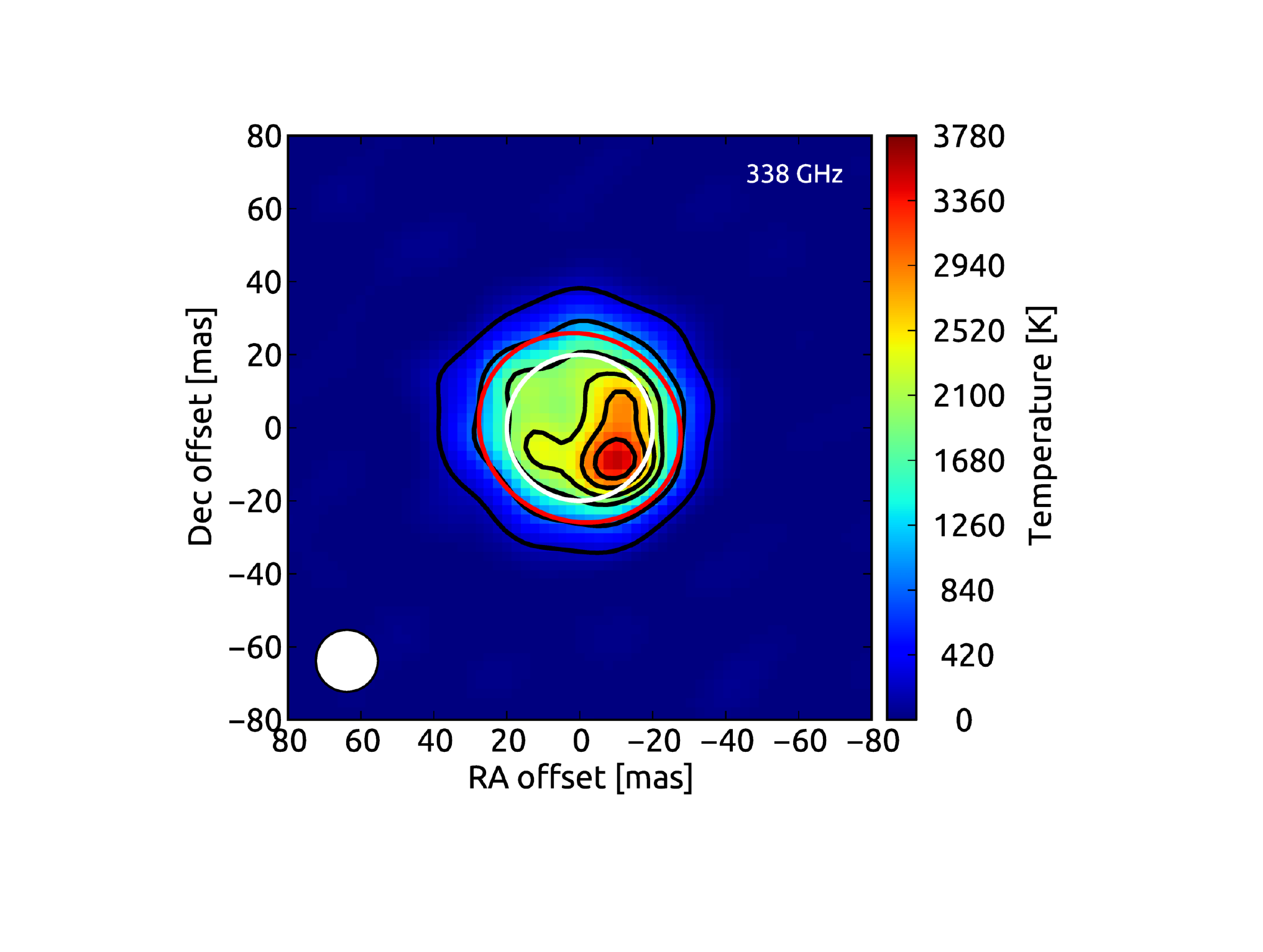} 
\caption{Brightness temperature map of the AGB star W~Hya observed
  with ALMA at 338 GHz (\cite[Vlemmings et al. 2017]{Vlemmings2017}). The red ellipse indicates
the size of the stellar disk at 338~GHz while the white circles
indicates the size of the optical photosphere. The clear hotspot is
unresolved and it brightness temperature in the map is a lower
limit. From size measurements we can constrain the true brightness
temperature to be $>50.000$~K, which could be a sign of shock
interaction or magnetic activity.}
\label{fig2}
\end{center}
\end{figure}

\section{Indirect tracers and origin of the magnetic field}

The origin of AGB magnetic fields is unclear and might require an
extra source of angular momentum to maintain a stellar dynamo. This
however depends strongly on the magnetic coupling throughout the star
itself. If a sufficiently strong magnetic field persist at the AGB
stellar surface, it might be possible to detect signs of magnetic
activity. Recently, it has been shown that the majority of the AGB
stars are UV-emitters (\cite[Montez et al. 2017]{Montez17}) which could
be a sign of (magnetic) activity. Similarly, recent observation of the
surface of the AGB star W Hya show high brightness temperature
hotspots (\cite[Fig~\ref{fig2}, Vlemmings et
al. 2017]{Vlemmings2017}). These spots can arise from strong shocks
but could also point to magnetic activity.

\begin{figure}
\centering
\includegraphics[width=13.5cm]{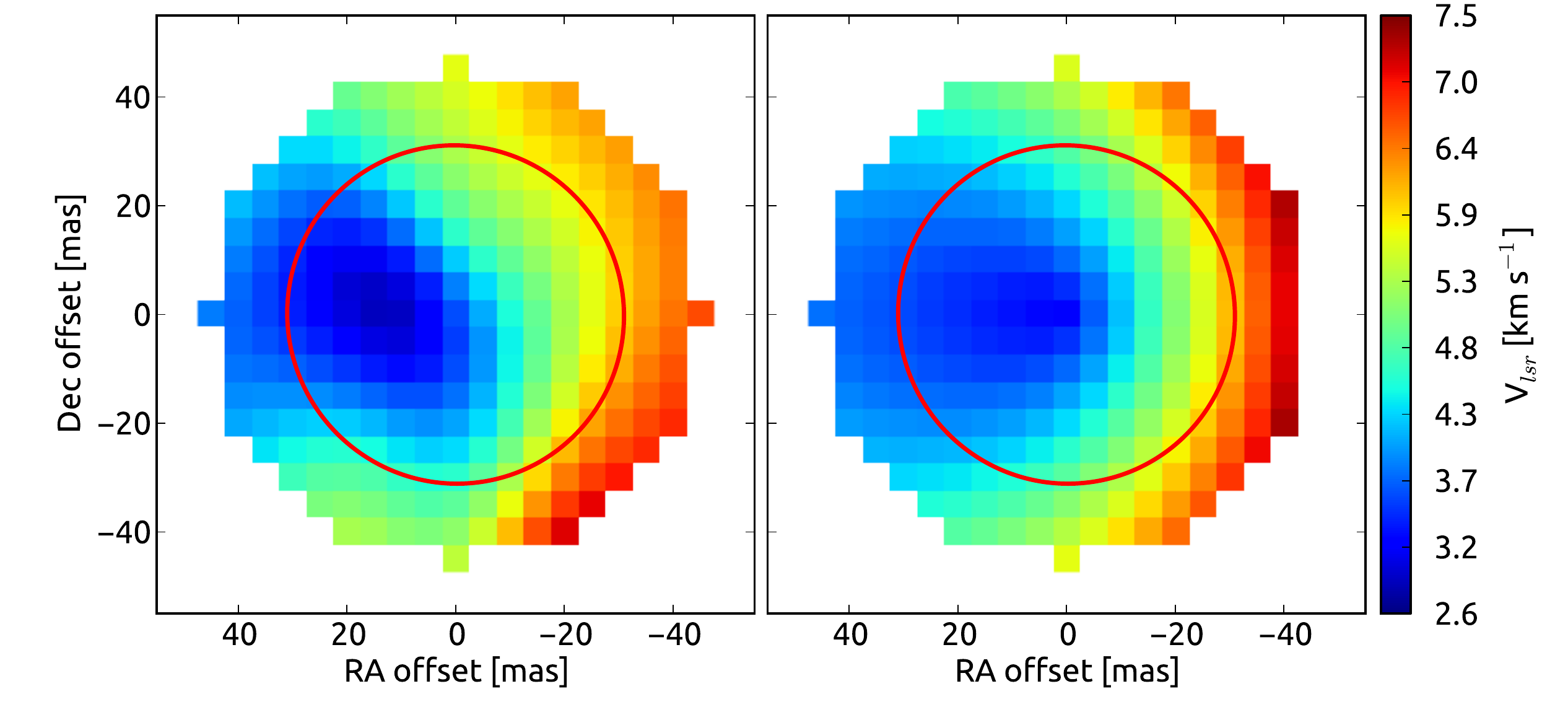}
\caption{From \cite[Vlemmings et al. 2018]{Vlemmings18b} {\it (left)}: Center velocity of the SiO $v=3, J=5-4$
emission line indicating the fast rotation in the envelope of R~Dot. The red ellipse indicates the measured size of
the star at $214$~GHz. {\it (right)}: The best fit model of
solid-body rotation including a small expansion velocity component.}
 \label{SiOfit}
\end{figure}

As previously noted, the angular momentum imparted by a stellar (or
sub-stellar) companion might be needed to maintain a stellar dynamo
that can generate the observed magnetic fields. However, rotation is
very difficult to measure for the extended AGB stars that are
undergoing pulsations and show large convective cells. Only very
recently has ALMA been able to measure the fast ($\sim1$~km~s$^{-1}$)
rotation of the AGB star R~Dor  (\cite[Fig~\ref{SiOfit}, Vlemmings et
al. 2018]{Vlemmings2018}). As the rotation is almost two orders
of magnitude larger than otherwise expected, it is a likely sign of
interaction with an hitherto unknown companion. Unfortunately, no
magnetic field observations exist yet for R~Dor and it it is thus not
yet possible to establish a link between the generation of a magnetic
field and the fast rotation.

\section{Conclusions}

Magnetic fields are ubiquitous around AGB and post-AGB stars, and
several observations indicate a link between the magnetic field and
the collimated outflows found in pre-PNe. Additionally, indirect
observations of hotspots and UV-emission might point to magnetic
activity on the surface of AGB stars. However, it is only now possible
to start probing the morphology of the magnetic field in AGB envelopes
and to finally determine the role of magnetism around evolved stars.

\begin{discussion}

\discuss{De Marco}{There seem to be too many AGB with B-fields to be justified by a close-by companion. So, are you saying that there \textit{must} be an alternative scenario to the binary scenario?}

\discuss{Vlemmings}{Yes. Although the sample can still be considered
  small, magnetic fields appear to be present in all studies sources
  with extrapolated surface field strength of a few Gauss. Certainly
  these sources do not all have close-by \textit{stellar} companions.}

\end{discussion}

\end{document}